\documentclass[pra,twocolumn,showpacs,superscriptaddress,eqsecnum]{revtex4}

\usepackage{amsfonts,amssymb,amsmath,bm,graphicx,times}

\newcommand{\op}[1]{\hat{#1}}
\newcommand{\Tr}{\mathop{\mathrm{Tr}}\nolimits}
\newcommand{\tr}{\mathop{\mathrm{tr}}\nolimits}
\newcommand{\prim}{\sigma}

\newcommand{\Gal}[1]{\mathrm{GF}(#1)}

\begin{document}

\title{Discrete coherent and squeezed states of many-qudit systems}

\author{Andrei~B.~Klimov}
\affiliation{Departamento de F\'{\i}sica, Universidad de Guadalajara, 
44420~Guadalajara, Jalisco, Mexico}

\author{Carlos~Mu\~{n}oz}
\affiliation{Departamento de F\'{\i}sica, Universidad de Guadalajara, 
44420~Guadalajara, Jalisco, Mexico}

\author{Luis~L.~S\'anchez-Soto}
\affiliation{Departamento de \'{O}ptica, Facultad de F\'{\i}sica, 
Universidad Complutense, 28040~Madrid, Spain}
\date{\today}

\begin{abstract}
  We consider the phase space for a system of $n$ identical qudits
  (each one of dimension $d$, with $d$ a primer number) as a grid of
  $d^{n} \times d^{n}$ points and use the finite field $\Gal{d^{n}}$
  to label the corresponding axes. The associated displacement
  operators permit to define $s$-parametrized quasidistribution
  functions in this grid, with properties analogous to their
  continuous counterparts. These displacements allow also for the
  construction of finite coherent states, once a fiducial state is fixed.  
  We take this reference as one eigenstate of the discrete Fourier transform
  and study the factorization properties of the resulting coherent states. 
  We extend these ideas to include discrete squeezed states, and show their 
  intriguing relation with entangled states between different qudits.
\end{abstract}

\maketitle

\section{Introduction}

The concept of phase-space representation of quantum mechanics,
introduced in the pioneering works of Weyl~\cite{Weyl:1928},
Wigner~\cite{Wigner:1932}, and Moyal~\cite{Moyal:1949}, is a very
useful and enlightening approach that sheds light on the
correspondence between quantum and classical worlds.

Numerous applications of the phase-space methods to physical 
problems have been extensively discussed in the last
decades~\cite{Lee:1995,Schroek:1996,Schleich:2001,QMPS:2005}. 
However, much of this subject is usually illustrated in terms of 
continuous variables, most often position and momentum.

For discrete systems, or qudits in the modern parlance of quantum
information, things are less straightforward. Since the dynamical
symmetry group for a qudit is SU($d$), one may be tempted to interpret
its associated phase space as a generalized Bloch
sphere~\cite{Kimura:2003,Schirmer:2004}, which is supported by the
rigorous construction of Kostant~\cite{Kostant:1970} and
Kirilov~\cite{Kirillov:1976} in terms of coadjoint orbits.  Even if
this picture is quite popular, especially when applied to qubits, one
can rightly argue that there is a lot of information redundancy there
and that the phase space should be a grid of points, as one could
expect for a truly discrete system.

Indeed, apart from some noteworthy
exceptions~\cite{Hannay:1980,Leonhardt:1995,Miquel:2002}, nowadays
there is a wide consensus in picturing the phase space for a qudit as
a $d \times d$ grid. This can be traced back to the elegant approach
proposed by
Schwinger~\cite{Schwinger:1960a,Schwinger:1960b,Schwinger:1960c}, who
clearly recognized that the expansion of arbitrary operators in terms
of certain operator basis was the crucial mathematical concept in
setting such a grid. These ideas have been rediscovered and developed
further by several authors~\cite{Buot:1973,Cohendet:1988, Kasperkovitz:1994,Opatrny:1995,Rivas:1999,Mukunda:2004,Chaturvedi:2006},
although the contributions of
Wootters~\cite{Wootters:1986,Wootters:1989,Wootters:2006,Wootters:2007}
and Galetti and
coworkers~\cite{Galetti:1988,Galetti:1992,Galetti:1995ly} are worth
stressing.

To equip this grid with properties analogous to the geometry of an
ordinary plane, it turns out
essential~\cite{Klimov:2005,Klimov:2007,Klimov:2009bk} to label the
axes in terms of the elements of a finite field $\Gal{d}$ with $d$
elements. It is well known that such a field exist only when the
dimension is a prime or a power of a prime~\cite{Lidl:1986}. This, of
course, gives a special role to qudits in prime dimensions, but also
is ideally suited to deal with systems of $n$ of these qudits.

Once the natural arena is properly established, the next question is
how to represent states (and operators) on phase space.  This is
done through quasidistribution functions, which allow for the
calculation of quantum averages in a way that exactly parallels
classical mechanics. There are, however, important differences with
respect to a classical description: they come from the noncommuting
nature of conjugate quantities (like position and momentum), which
precludes their simultaneous precise measurement and, therefore,
imposes a fundamental limit on the accuracy with which we can
determine a point in phase space. As a distinctive consequence of
this, there is no unique rule by which we can associate a classical
phase-space variable to a quantum operator.  Therefore, depending on
the operator ordering, various quasidistributions can be defined.  For
continuous variables, the best known are the
Glauber-Sudarshan $P$ function~\cite{Glauber:1963,Sudarshan:1963}, 
the Husimi $Q$ function~\cite{Husimi:1940}, and the
Wigner $W$ function~\cite{Hillery:1984}, corresponding to normal, 
antinormal, and symmetric order, respectively, in the associated 
characteristic functions. In fact, all of them are special cases
of the $s$-parametrized quasidistributions introduced by
Cahill and Glauber~\cite{Cahill:1969}.

The problem of generalizing these quasidistributions (mainly the
Wigner function) to finite systems has also a long history.  Much of
the previous literature has focused on spin variables, trying to
represent spin states by continuous functions of angle variables.
This idea was initiated by Stratonovich~\cite{Stratonovich:1956},
Berezin~\cite{Berezin:1975}, and Agarwal~\cite{Agarwal:1981}.  The
resulting Wigner function, naturally related to the SU(2) dynamical
group, has been further studied by a number of
authors~\cite{Scully:1983,Cohen:1986,Varilly:1989,Heiss:2000}, has
been applied to some problems in quantum
optics~\cite{Dowling:1994,Benedict:1999} and extended to more general
groups~\cite{Brif:1998}.

However, these Wigner functions are not defined in a discrete phase
space.  A detailed review of possible solutions can be found in
Ref.~\cite{Bjork:2008}.  Perhaps, the most popular one is due to
Wootters and
coworkers~\cite{Wootters:1987,Gibbons:2004,Wootters:2004}, which
imposes a structure by assigning a quantum state to each line in phase
space. Any good assignment of quantum states to lines is called a
``quantum net'' and can be used to define a discrete Wigner
function. In this paper, we show how to introduce a set of
$s$-parametrized functions, in close correspondence with the
continuous case. We emphasize that, although some interesting work has
been done in this direction by using a mod~$d$
invariance~\cite{Ruzzi:2005sd,Marchiolli:2005rm}, our approach works
quite well for many-qudit systems.

Another essential ingredient in any phase-space description is the
notion of coherent states~\cite{Klauder:1999}. This is firmly
established for continuous variables and can easily extended for other
dynamical symmetry groups~\cite{Perelomov:1986}. However, for discrete
systems we have again a big gap waiting to be filled. The reason for
this can be traced back to the fact that, as brightly pointed out in
Ref.~\cite{Ruzzi:2006}, in the continuum we have one, and only one,
harmonic oscillator, while in the discrete there are a lot of
candidates for that role, each one surely with its virtues, but surely
no undisputed champion.

The strategy we adopt to deal with this problem is to look for
eigenstates of the discrete Fourier transform~\cite{Galetti:1996}. 
For continuous variables,  they have a very distinguishable behavior that is 
at the basis of the remarkable properties of coherent states. We explore this 
approach, getting a strikingly simple family of discrete coherent states
(even for many qudits) fulfilling all the requirements.

To put the icing on the cake, we also extend the notion of squeezed
states for these systems~\cite{Marchiolli:2007}. For a single qudit,
the resulting states have nice and expected properties. However, when
they really appear as highly interesting is for multipartite systems,
since they present an intriguing relation with entanglement.

In short, the program developed in this paper can be seen as a handy
toolbox for the phase-space analysis of many-qudit systems, which
should be of interest to a large interdisciplinary community working
in these topics.

\section{Phase space for continuous variables}

\label{sec:qpWig}

In this Section we briefly recall the relevant structures needed to
set up a phase-space description of Cartesian quantum mechanics.  This
will facilitate comparison with the discrete case later on.  For
simplicity, we choose one degree of freedom only, so the associated
phase space is the plane $\mathbb{R}^2$.

The relevant observables are the Hermitian coordinate and momentum
operators $\op{q}$ and $\op{p}$, with canonical commutation relation
(with $\hbar =1$ throughout)
\begin{equation} 
  [ \op{q},\op{p} ]= i \, \op{\openone} \, , 
\label{eq:HWcom}
\end{equation}
so that they are the generators of the Heisenberg-Weyl
algebra. Ubiquitous and profound, this algebra has become the hallmark
of noncommutativity in quantum theory. To avoid technical problems
with the unbounded operators $\op{q}$ and $\op{p}$, it is convenient
to work with their unitary counterparts~\cite{Putnam:1967}
\begin{equation}
  \op{U}(q) = \exp (-iq \,\op{p}) \, ,
  \qquad 
  \op{V}(p)=\exp (ip\,\op{q}) \, ,
  \label{UV}
\end{equation}
which generate translations in position and momentum, respectively. 
The commutation relations are then expressed in the Weyl form
\begin{equation}
  \op{V}(p) \op{U}(q) = e^{iqp} \, \op{U}(q) \op{V}(p) \, .  
  \label{eq:Weyl}
\end{equation}
Their infinitesimal form immediately gives (\ref{eq:HWcom}), but
(\ref{eq:Weyl}) is more useful in many instances.

In terms of $\op{U}$ and $\op{V}$ a displacement operator can be
introduced as
\begin{equation}
  \label{eq:HWDisp1} 
  \op{D} (q,p) = e^{- i q p/2} \,  \op{U} (p) \op{V}(q) \, ,
\end{equation}
which usually is presented in the entangled form $\op{D} (q,p) =
\exp[i(p \op{q} - q \op{p})]$. However, this cannot be done in more
general situations.

The $\op{D} (q, p)$ form a complete orthonormal set (in the trace
sense) in the space of operators acting on $\mathcal{H}$ (the Hilbert
space of square integrable functions on $\mathbb{R}$). The unitarity
imposes that $\op{D}^\dagger (q, p) = \op{D} (-q, - p)$, and
$\op{D}(0,0) = \op{\openone}$.  In addition, they obey the simple
composition law
\begin{equation}
  \label{eq:com} 
  \op{D} (\op{q}_{1}, \op{p}_{1}) \op{D} (\op{q}_{2}, \op{p}_{2}) = 
  e^{i (p_{1} q_{2} - q_{1} p_{2})/2} \, 
  \op{D} (\op{q}_{1} + \op{q}_{2}, \op{q}_{2} + \op{p}_{2}) \, .
\end{equation}

The displacement operators constitute a basic element for the
notion of coherent states. Indeed, if we choose a fixed normalized
reference state $ |\psi_{0}\rangle $, we can define these coherent 
states as~\cite{Perelomov:1986}
\begin{equation}
  | q, p \rangle = \op{D} ( q, p) \, | \psi_{0} \rangle \, ,  
  \label{eq:defCS}
\end{equation}
so they are parametrized by phase-space points. These states
have a number of remarkable properties, inherited from those of
$\op{D} (q, p)$. In particular, $\op{D} (q, p)$ transforms any coherent
state in another coherent state:
\begin{equation}
  \op{D} ( \op{q}_{1}, \op{p}_{1} ) \, | q_{2}, p_{2} \rangle
  = e^{i (p_{1} q_{2} - q_{1} p_{2})/2} \,
 | q_{1} + q_{2}, q_{2} + p_{2} \rangle \, .
  \label{eq:comcoh}
\end{equation}
We need to determine the fiducial vector $| \psi_{0}\rangle $. The
standard choice is to take it as the vacuum $|0\rangle $. This has
quite a number of relevant properties, but the one we want to stress
for what follows is that $| 0 \rangle $ is an eigenstate of the
Fourier transform (as they are all the Fock states)~\cite{Peres:1993}, 
and so is the Gaussian
\begin{equation}
  \psi_{0} (q) = \langle q | 0 \rangle =
  \frac{1}{\pi^{1/4}} \, \exp ( - q^{2}/2) \, ,
  \label{eq:Gauss}
\end{equation}
in appropriate units. In addition, this wavefunction represents a
minimum uncertainty state, namely
\begin{equation}
  ( \Delta q )^{2} \, ( \Delta p)^{2} = \frac{1}{4} \,,  
  \label{eq:MUS}
\end{equation}
where $(\Delta q)^{2}$ and $(\Delta p)^{2}$ are the corresponding
variances.

Our next task is to map the density matrix $\op{\varrho}$ into a
function defined on $\mathbb{R}^{2}$. There exists a whole class of
these quasidistribution functions, related to different orderings of
$\op{q}$ and $\op{p}$. The corresponding mappings are generated by an
$s$-ordered kernel
\begin{equation}
  W_{\op{\varrho}}^{(s)} ( q, p ) = 
  \Tr [ \op{\varrho} \, \op{w}^{(s)} (q, p) ] \, ,  
  \label{eq:Wigcan}
\end{equation}
where $\op{w}^{(s)}$ is the double Fourier transform of the
displacement operator with a weight fixed by the operator ordering
\begin{eqnarray}
  \op{w}^{(s)} (q , p) & =  & \frac{1}{(2\pi)^{2}} 
  \int_{\mathbb{R}^{2}}  \exp [-i (p q^{\prime} - q p^{\prime} ) ] \, 
  \op{D} ( q^{\prime}, p^{\prime} ) \nonumber \\
  & \times &
  \langle \psi_{0} | \op{D} (q^{\prime}, p^{\prime} ) | \psi_{0} \rangle^{-s}
  \,  dq^{\prime} dp^{\prime} \, ,  
  \label{eq:HWkernelDef}
\end{eqnarray}
and $s \in [-1 , 1]$. The mapping is invertible, so that
\begin{equation}
  \op{\varrho} = \frac{1}{(2\pi)^{2}} \int_{\mathbb{R}^{2}} 
  \op{w}^{(-s)} (q , p) \, W^{(s)}(q,p) \, dq dp \, .
\end{equation}

The Hermitian kernels $\op{w}^{(s)} (q, p)$ are also a complete
trace-orthonormal set and they transform properly under displacements
\begin{equation}
  \op{w}^{(s)} ( q, p ) = 
  \op{D} ( q, p) \, \op{w}^{(s)}(0,0) \,\op{D}^{\dagger} ( q, p) \, .  
  \label{eq:HWKernelDisp}
\end{equation}
The symmetric ordering ($s = 0$) corresponds to the Wigner function
$W (q, p)$ and the associated kernel $\op{w}^{(0)} (0,0)$ is just $ 2
\op{\mathcal{P}}$, where $\op{\mathcal{P}}$ is the parity operator.
For the antinormal ordering ($s=-1$), which corresponds to the Husimi 
$Q$ function, $\op{w}^{(0)}(0,0) = | 0 \rangle \langle 0 |/\pi$.

The quasidistribution functions (\ref{eq:Wigcan}) fulfill all the
basic properties required for any good probabilistic description.
First, due to the Hermiticity of $\op{w}^{(s)}(q,p)$, they are real 
for Hermitian operators. Second, on integrating $W^{(s)}(q,p)$ over 
one variable, the probability distribution of the conjugate variable 
is reproduced. And finally, $W^{(s)}(q,p)$ is covariant, which means 
that for the displaced state $\op{\varrho}^{\prime} = 
\op{D} (q_{0}, p_{0}) \, \op{\varrho} \, \op{D}^{\dagger} (q_{0}, p_{0})$, 
one has
\begin{equation}
  W_{\op{\varrho}^\prime}^{(s)} (q, p) = 
  W_{\op{\varrho}}^{(s)} (q - q_{0}, p - p_{0}) \, ,  
  \label{eq:HWProps3}
\end{equation}
so that these functions follow displacements rigidly without changing
their form, reflecting the fact that physics should not depend on a
certain choice of the origin.

\section{Single qudit}

\subsection{Discrete phase space}

We consider a system living in a Hilbert space $\mathcal{H}_{d}$,
whose dimension $d$ is assumed from now on to be a prime number.  
We choose a computational basis $| \ell \rangle $ in $\mathcal{H}_{d}$
($\ell = 0, \ldots, d-1$) which we arbitrarily interpret as the
``position'' basis, with periodic boundary conditions $| \ell + d
\rangle = | \ell \rangle$. The conjugate ``momentum'' basis can 
be introduced by means of the discrete Fourier
transform~\cite{Vourdas:2004}, that is
\begin{equation} 
\label{eq:conjBas} 
| \tilde{\ell} \rangle  = \op{\mathcal{F}} \, | \ell \rangle \, ,
\end{equation}
where
\begin{equation}
  \label{FT1} 
  \op{\mathcal{F}} = \frac{1}{\sqrt{d}}
  \sum_{\ell , \ell^\prime = 0}^{d-1} \omega(\ell \, \ell^{\prime}) \,
  |\ell \rangle \langle \ell^{\prime}| \, ,
\end{equation}
and we use the notation
\begin{equation}
  \omega ( \ell ) = \omega^{\ell} = \exp (i 2\pi \ell/d) \, ,
\end{equation}
$\omega = \exp( i 2\pi/d)$ being a $d$th root of the unity.
Whenever we do not specify the ranges in a sum, we understand
the index running all its natural domain.

Once we have position and momentum basis, the phase space turns 
out to be a periodic $d \times d$ grid of points; i.e., the torus
$\mathbb{Z}_{d} \times \mathbb{Z}_{d}$, where $\mathbb{Z}_{d}$ 
is the field of the integer numbers modulo $d$. 

Mimicking the approach in Sec.~\ref{sec:qpWig}, we introduce the
operators $\op{U}$ and $\op{V}$, which generate finite translations 
in position and momentum, respectively.  In fact, $\op{U}$ generates
cyclic shifts in the position basis, while $ \op{V}$ is diagonal
\begin{equation}
  \label{CC}
  \op{U}^{n} | \ell \rangle = | \ell + n \rangle \, , 
  \qquad 
  \op{V}^{m} | \ell \rangle = \omega(m \ell) \, | \ell \rangle \, , 
\end{equation}
where addition and multiplication must be understood
modulo~$d$. Conversely, $\op{U}$ is diagonal in the momentum basis and
$\op{V}$ acts as a shift, which is reflected also by the fact that
\begin{equation}
  \label{FT2} 
  \op{V} = \mathcal{F} \, \op{U} \,  \mathcal{F}^\dagger \, ,
\end{equation}
much in the spirit of the standard way of looking at complementary
variables in the infinite-dimensional Hilbert space: the position and
momentum eigenstates are Fourier transform one of the other. Note that
the operators $ \op{U}$ and $\op{V}$ are generalizations of the Pauli
matrices $\sigma_{x}$ and $\sigma_{z}$, so many authors use the
notation $\op{X}$ and $\op{Z}$ for them.

One can directly check the identity
\begin{equation}
  \label{eq:Weyldis}
  \op{V}^{m} \op{U}^{n} = \omega (m n) \, \op{U}^{n} \op{V}^{m} \, ,
\end{equation}
which is the finite-dimensional version of the Weyl form of the
commutation relations and show that they obey a generalized Clifford
algebra~\cite{Chuang:2000}.

One may be tempted to define discrete position and momentum
operators. A possible way to achieve this is to
write~\cite{Vourdas:2005}
\begin{equation}
  \op{U} = \exp (- i 2 \pi  \op{P} / d ) \, , 
  \qquad 
  \op{V} = \exp ( i 2 \pi  \op{Q} / d ) \, ,
\end{equation}
with
\begin{equation}
  \op{Q} = \sum_{\ell} \ell \, | \ell \rangle \langle \ell | \, , 
  \qquad 
  \op{P} = \sum_{\tilde{\ell}} \tilde{\ell} \, 
  | \tilde{\ell} \rangle \langle  \tilde{\ell} | \, .
\end{equation}
However, for finite quantum systems the Heisenberg-Weyl group is
discrete, there is no Lie algebra (that is, there are no infinitesimal
displacements) and the role of position and momentum is limited. For
this reason our formalism is mainly based on the operators $\op{U}$
and $\op{V}$.

Next we introduce the displacement operators
\begin{equation}
  \op{D}(m,n)= \phi (m,n) \, \op{U}^{n}\op{V}^{m} \, ,  
  \label{eq:desp}
\end{equation}
where $\phi (m,n)$ is a phase required to avoid plugging extra factors
when acting with $\op{D}$. The conditions of unitarity and periodicity
restrict the possible values of $\phi$. One relevant choice (for $d>2$)
that have been analyzed in the literature is~\cite{Bjork:2008}
\begin{equation}
  \phi (m, n) = \omega( 2^{-1} \,m n) \, ,  
  \label{phi1}
\end{equation}
where $2^{-1}$ is the multiplicative inverse of $2$ in
$\mathbb{Z}_{d}$. For qubits, $\phi(m, n)$ may be taken as $ \phi
(m,n) = \pm i^{mn}$.

Without entering in technical details, this choice guarantees all the
good properties, in particular the analogous to Eq.~(\ref{eq:com}):
\begin{eqnarray}
  \label{eq:comsq}
  & \op{D}(m_{1},n_{1})\op{D}(m_{2},n_{2})   =   
  \omega [ 2^{-1}(m_{1}n_{2}-m_{2}n_{1})] &  \nonumber  \\
  & \times  \op{D}(m_{1}+m_{2},n_{1}+n_{2})  \, , &
\end{eqnarray}
and the following relation
\begin{equation}
  \frac{1}{d} \sum_{m,n} \op{D}(m,n) = \op{\mathcal{P}} \, ,  
  \label{eq:parity}
\end{equation}
where $\op{\mathcal{P}}$ is the parity operator $\op{\mathcal{P}} |
\ell \rangle = | - \ell \rangle $ modulo $d$. Physically, this is the
basis for translational covariance and this also means that $\op{D}
(m, n)$ translates the standard basis states cyclically $m$ places in
one direction and $n$ places in the orthogonal one, as one would
expect from a displacement operator.

\subsection{Coherent states}

Once a proper displacement operator has been settled, the coherent
states for a single qudit can be defined as
\begin{equation}
  \label{eq:defcoh} 
   | m, n \rangle = \op{D} (m, n) \, | \psi_{0} \rangle \, ,
\end{equation}
where $| \psi_{0} \rangle$ is again a reference state. These states
are also labeled by points of the discrete phase space, as it should be.

A possible choice~\cite{Saraceno:1990,Paz:2004} is to use for 
$| \psi_{0} \rangle$  the ground state of the Harper Hamiltonian~\cite{Harper:1955}
\begin{equation}
  \label{eq:Harper}
  \op{H} = 2 - \frac{\op{U} + \op{U}^\dagger}{2} - 
  \frac{\op{V} + \op{V}^\dagger}{2} \, ,
\end{equation}
which is considered as the discrete counterpart of the harmonic
oscillator with the proper periodicity conditions. While 
such a replacement is interesting, it is by no means unique.

We prefer to take a different route, pioneered by Galetti and
coworkers~\cite{Galetti:1996}.  We use again as a guide
the analogy with the continuous case and look for eigenstates $ | f
\rangle $ of the discrete Fourier transform, which play the role of
Fock states for our problem and are determined by
\begin{equation}
  \langle \ell | \op{\mathcal{F}} | f \rangle =
  i^{\ell} \,  \langle \ell | f \rangle \, .
\end{equation}
Obviously, the fact that $\op{\mathcal{F}}^{4} = \op{\openone}$
implies that it has four eigenvalues: 1, $-1$, $i$, and $-i$. The
solutions of this equation were fully studied by
Mehta~\cite{Mehta:1987} (see also Ruzzi~\cite{Ruzzi:2006}). Taking $|
\psi_{0} \rangle $ as the ``ground'' state (i.e., $\ell =0$) one gets
\begin{equation}
  | \psi_{0} \rangle = \frac{1}{\sqrt{C}} \sum_{k \in \mathbb{Z}}
  \sum_{\ell} \omega (k \ell )  \, e^{-\frac{\pi}{d}k^{2}} \, 
  | \ell \rangle \, ,  
  \label{0_cs}
\end{equation}
and the normalization constant $C$ is given by
\begin{equation}
  C = \sum_{k \in \mathbb{Z}} e^{-\frac{2\pi}{d}k^{2}} = 
  \vartheta_{3} \left ( 0 \bigl | e^{-\frac{2\pi}{d}} \right ) \, ,
\end{equation}
$\vartheta _{3}$ being the third Jacobi function~\cite{Mumford:1983}.
Note in passing that this fiducial state can be alternatively
represented as
\begin{equation}
  | \psi_{0} \rangle = \frac{1}{\sqrt{C}} \sum_{\ell} 
  \vartheta_{3} \left ( \frac{\pi \ell}{d} \bigl | 
    e^{- \frac{\pi}{d}} \right) | \ell \rangle \, .
  \label{eq:theta3}
\end{equation}
The appearance of the Jacobi function in the present context
can be directly understood by realizing that this function 
is a periodic eigenstate of the discrete Fourier
operator with eigenvalue $+1$ and period $d$. In addition, it plays
the role of the Gaussian for periodic variables, which makes this
approach even more appealing~\cite{Rehacek:2008}.

We also observe that $| \psi_{0} \rangle$ satisfies a ``parity''
condition: if we write it as $| \psi_{0} \rangle = \sum_{\ell}
c_{\ell} \, | \ell \rangle$, then $c_{\ell} = c_{- \ell}$. This
guarantees that the average values of $\op{U}$ and $\op{V}$ in 
$| \psi_{0} \rangle$ are the same: $\langle \psi_{0} | \op{U} | \psi_{0}
\rangle = \langle \psi_{0} | \op{V} |\psi_{0} \rangle$.

The coherent states (\ref{eq:defcoh}) have properties fully analogous 
to the standard ones for continuous variables, as one can check with
little effort. 

The Harper Hamiltonian commutes with the Fourier operator
$[\op{\mathcal{F}},\op{H}]=0$. In fact, the state (\ref{0_cs}) is an
approximate eigenstate of (\ref{eq:Harper}) in the high-dimensional
limit
\begin{equation}
  \op{H} | \psi_{0} \rangle \simeq  \left ( \frac{\pi}{d} -
    \frac{\pi^{2}}{2d^{2}} + \frac{\pi^{3}}{6d^{3}} \right) 
  | \psi_{0} \rangle \, , 
  \qquad d \gg 1 \, ,  
  \label{HS_07}
\end{equation}
which provides another argument for its use as a reference.

Finally, according to the recent results in Refs.~\cite{Forbes:2003}
and \cite{Massar:2008}, the following uncertainty relation holds
\begin{equation}
  \label{eq:DUDV}
  (\Delta U)^{2} \, (\Delta V)^{2} \ge \frac{\pi^{2}}{d^{2}} \, ,
\end{equation}
where $(\Delta U)^{2} = 1 - | \langle \psi | \op{U} | \psi \rangle
|^{2}$ [and an analogous expression for $(\Delta V)^{2}$] denotes the
circular dispersion, which is the natural generalization of variance
for unitary operators. One can check that $| \psi_{0} \rangle$
saturates this inequality, confirming that it is also a minimum
uncertainty state.
 
\subsection{Quasidistribution functions}

The displacement operators lead us to introduce a Hermitian
$s$-ordered kernel
\begin{eqnarray}
  \op{w}^{(s)} (m, n) & =  & \frac{1}{d} \sum_{k,l} \omega (nk - ml) \, 
  \op{D}(m,n) \nonumber \\
  & \times &  \langle \psi_{0} | \op{D} (m, n) | \psi_{0} \rangle^{-s} \, ,  
  \label{w_s_d}
\end{eqnarray} 
which, as $\op{w}^{(s)} (q, p)$ in Eq.~(\ref{eq:HWkernelDef}), appears 
as a double Fourier transform of $\op{D}$ with a weight determined by the 
operator ordering. However, here the parameter $s$ takes only discrete 
values ($s = -1, 0, 1$).  

These kernels are normalized and covariant under transformations of the
generalized Pauli group
\begin{equation}
  \op{D} (m, n) \, \op{w}^{(s)} (k, l) \,\op{D}^{\dagger} (m, n) = 
  \op{w}^{(s)} (k + m, l + n) \, .
  \label{eq:cov}
\end{equation}
They can be then conveniently represented as
\begin{equation}
  \op{w}^{(s)} (m, n) = \op{D} (m, n) \,\op{w}^{(s)} (0,0) \,
  \op{D}^{\dagger} (m, n) \, ,
  \label{eq:par}
\end{equation} 
where, according to Eq.~(\ref{eq:parity}), $\op{w}^{(s)} (0,0)$ coincides 
with the parity for $s = 0$ ($d \neq 2$), as in the continuous case.

The  $s$-ordered quasidistribution functions $W^{(s)}_{\op{\varrho}}$ 
are generated through the mapping
\begin{equation}
  \label{eq:eds}
  W^{(s)}_{\op{\varrho}} ( m, n) = \Tr [\op{\varrho} \, 
  \op{w}^{(s)} (m, n) ] \, ,
\end{equation}
which is invertible, so that
\begin{equation}
  \op{\varrho} = \frac{1}{d} \sum_{m,n}  \op{w}^{-(s)} (m,n) \, 
  W^{(s)} (m, n) \, .
\end{equation}
These functions fulfill all the basic properties required for the
probabilistic description we are looking for. Let us apply them to the
reference state $|\psi_{0} \rangle$ (notice that any other coherent
state is just a displaced copy of this one). The corresponding Wigner
function ($s=0$) can be obtained after some algebra. We omit the
details and merely quote the final result:
\begin{eqnarray}
  \label{wfcs_1}
  W_{|\psi_{0} \rangle} ( m , n ) & =  & 
  \frac{d}{C} \sum_{k}  \sum_{p,q \in \mathbb{Z}} 
  \omega  [( 2k - 1 - 2m ) n ]  \nonumber \\
  & \times & \exp [ - k + 2 m + q d - (d-1)/2]^{2}    \nonumber \\
  & \times &   \exp  [ -(\pi / d)  ( k + p d - d/2 )^{2}]   \, ,
\end{eqnarray}
which, in the limit $d\gg 1$, can be approximated by the compact
expression
\begin{eqnarray}
  W^{(0)}_{|\psi_{0} \rangle} ( m , n ) & \simeq &  
  \frac{\sqrt{2}}{d^{3/2}} \sum_{k,l} (-1)^{kl}  \, 
  \omega (mk-nl) \nonumber \\
  & \times & \exp [ - \pi (k^{2}+l^{2})/(2d) ] \, . 
\end{eqnarray}

%%%%%%%%%%%%%%%%%%%%%%%%%%%%%%%%%%%%%%%%%%%%%%%%%
\begin{figure}
  \includegraphics[width=0.85\columnwidth]{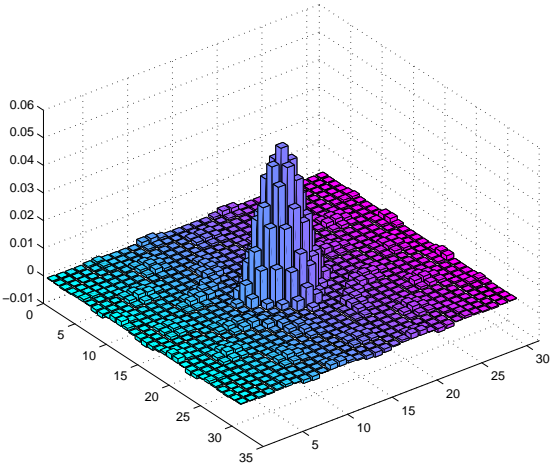}
  \caption{$Q$ function (\ref{eq:Qf0}) for the reference state $|
    \psi_{0} \rangle$, which plays the role of vacuum for continuous
    states.}
  \label{Fig1}
\end{figure}
%%%%%%%%%%%%%%%%%%%%%%%%%%%%%%%%%%%%%%%%%%%%%%%%

The $Q$ function ($s = -1$) for the same state $| \psi_{0}\rangle$
reduces to
\begin{equation}
  Q_{|\psi_{0} \rangle} = | \langle \psi_{0} | \op{D}(m,n) | \psi_{0} \rangle |^{2} \, ,
  \label{eq:Qf0}
\end{equation}
which exhibits the additional interesting symmetry
\begin{equation}
  Q_{|\psi_{0} \rangle} (m, n) = Q_{|\psi_{0} \rangle} (- n, m ) \, .
\end{equation}
In Fig.~\ref{Fig1} we have plotted this $Q$ function for a
31-dimensional system.  The aspect of the figure confirms the issues
one expects from a fairly localized Gaussian state.

\section{Many qudits}

\subsection{Discrete phase space}

Next, we consider a system of $n$ identical qudits, living in the
Hilbert space $\mathcal{H}_{d^{n}}$. Instead of natural numbers, it is
convenient to use elements of the finite field $\Gal{d^{n}}$ to label
states: in this way we can almost directly translate all the
properties studied before for a single qudit and we can endow the
phase-space with many of the geometrical properties of the ordinary
plane~\cite{Gibbons:2004}. In the Appendix we briefly summarize the
basic notions of finite fields needed to proceed.

We denote by $|\lambda \rangle$ [from here on, Greek letters will
label elements in the field $\Gal{d^n}$] an orthonormal basis
in the Hilbert space of the system. Operationally, the elements of the
basis can be labeled by powers of a primitive element using, for
instance, the polynomial or the normal basis.

The generators of the Pauli group act now as
\begin{equation}
  \label{eq:XZgf}
  \op{U}_{\nu} | \lambda \rangle = 
  | \lambda + \nu \rangle \, , 
  \qquad 
  \op{V} _{\mu} | \lambda \rangle = 
  \chi ( \mu \lambda ) \, | \lambda \rangle \, ,
\end{equation}
where $\chi (\lambda )$ is an additive character (defined in the
Appendix) and  the Weyl form of the commutation relations reads as
\begin{equation}
  \op{V}_{\mu} \op{U}_{\nu} = \chi ( \mu \nu ) \,  
  \op{U}_{\nu} \op{V}_{\mu} \, .
\end{equation}

The finite Fourier transform~\cite{Vourdas:2007}
\begin{equation}
  \label{FTcomp}
  \op{\mathcal{F}}  = \frac{1}{\sqrt{d^{n}}} 
  \sum_{\lambda, \lambda^\prime} 
  \chi (\lambda \, \lambda^\prime) \, 
  | \lambda \rangle \langle \lambda^\prime |
\end{equation}
allows us to introduce the conjugate basis $| \op{\lambda} \rangle =
\op{\mathcal{F}} | \lambda \rangle$ and also we have
\begin{equation}
  \label{FT3} 
  \op{V}_{\mu} = 
  \op{\mathcal{F}} \,  \op{U}_{\mu} \, \op{\mathcal{F}}^\dagger \, .
\end{equation}
In this way, the  concepts delineated in the previous section can
be immediately generalized. For example, the displacement operators are 
\begin{equation}
  \op{D}(\mu ,\nu ) = \phi (\mu ,\nu ) \, \op{U}_{\nu} \op{V}_{\mu}  \, ,
  \label{eq:discom}
\end{equation}
where the phase $\phi (\mu ,\nu )$ must satisfy the conditions
\begin{equation}
  \label{eq:condphi}
  \phi ( \mu , \nu ) \, \phi^{\ast} (\mu ,\nu ) = 1 \, ,
  \qquad  
  \phi ( \mu , \nu ) \, \phi ( - \mu , - \nu ) = \chi ( - \mu \nu )  \, ,
\end{equation}
to guarantee the unitarity and orthogonality of $\op{D}$.  We also
impose $\phi (\mu ,0)=1$ and $\phi (0, \nu )=1$, which means that the
displacements along the ``position'' axis $\mu $ and the ``momentum'' 
axis $\nu $ are not associated with any phase.

For fields of odd characteristics one possible form of this phase is
\begin{equation}
  \label{eq:phiodd}
  \phi (\mu, \nu) = \chi (- 2^{-1} \, \mu \nu)  \, ,
\end{equation}
and we have then the same composition law as in Eq.~(\ref{eq:comsq}),
namely
\begin{eqnarray}
  \op{D} (\mu_{1}, \nu_{1}) \op{D} (\mu_{2}, \nu_{2}) & = & 
  \chi [ 2^{-1} (\mu_{1} \nu_{2} - \mu_{2} \nu_{1})]  \nonumber \\
  & \times & \op{D} (\mu_{1} + \mu_{2}, \nu_{1} + \nu_{2}) \, .
\end{eqnarray}

\subsection{Coherent states}

Given our previous discussion, it seems reasonable to extend the
coherent states~(\ref{eq:defcoh}) in the form
\begin{equation}
  | \mu ,\nu \rangle = \op{D}(\mu , \nu ) \, |\Psi_{0} \rangle \, ,
  \label{eq:defcoh_n}
\end{equation}
where $| \Psi_{0} \rangle$ is a reference state to be determined.  In
the continuous case, the extension of coherent states (\ref{eq:defCS})
to many degrees of freedom is straightforward: they are simply
obtained by taking the direct product of single-mode coherent
states. To reinterpret (\ref{eq:defcoh_n}) in the same spirit, one
needs first to map the abstract Hilbert space $\mathcal{H}_{d^n}$,
where the $n$-qudit system lives, into $n$ single-qudit Hilbert spaces
$\mathcal{H} _{d} \otimes \cdots \otimes \mathcal{H}_{d}$. This is
achieved by expanding any field element in a convenient basis $\{
\theta_{j} \}$ ($j = 1, \ldots, n$), so that
\begin{equation}
  \lambda = \sum_{j} \ell_{j} \, \theta_{j} \, ,
\end{equation}
where $\ell_{j} \in \mathbb{Z}_{d}$. Then, we can represent the states
as
\begin{equation}
  | \lambda \rangle = 
  | \ell_{1} \rangle \otimes \cdots \otimes | \ell_{n} \rangle =
  | \ell_{1}, \ldots, \ell_{n} \rangle \, ,
\end{equation} 
and the coefficients $\ell_{j}$ play the role of quantum numbers for
each qudit.

The use of the selfdual basis is especially advantageous, since only
then the basic operators (and the Fourier operator) factorize in terms
of single-qudit analogues
\begin{equation}
  \op{U}_{\nu} = \op{U}^{n_{1}} \otimes \cdots \otimes 
  \op{U}^{n_{n}} \, , 
  \qquad 
  \op{V}_{\mu} = \op{V}^{m_{1}} \otimes \cdots \otimes 
  \op{V}^{m_{n}} \, ,
\end{equation}
and the displacement operators factorize accordingly
\begin{equation}
  \op{D} (\mu, \nu) = \op{D}(m_{1}, n_{1}) \otimes \cdots \otimes 
  \op{D} (m_{n}, n_{n} ) \, ,
\end{equation}
where $m_{j}, n_{j} \in \mathbb{Z}_{d}$ are the coefficients of the
expansion of $\mu$ and $\nu$ in the basis, respectively. In consequence,
the eigenstates of the Fourier transform are direct product of
single-qudit eigenstates and we can write for the reference state
\begin{equation}
  | \Psi_{0} \rangle = \bigotimes_{j=1}^{n} |\psi_{0j} \rangle \, ,
  \label{0_csmult}
\end{equation}
where $|\psi_{0j}\rangle $ are of the form (\ref{0_cs}) for each qudit
(with $d > 2$).  For qubits, we have~\cite{Munoz:2009}
\begin{equation}
  | \Psi_{0} \rangle = \bigotimes_{j=1}^{n} 
  \frac{ ( | 0  \rangle + \xi | 1 \rangle )_{j}}
  {( 1 +  \xi^{2} )^{1/2}} \, , 
  \label{Psi_n2}
\end{equation}
with $\xi =\sqrt{2} - 1$.

Unfortunately, the selfdual basis can be constructed only if either
$d$ is even or both $n$ and $d$ are odd. This means that for such a
simple system as two qutrits, this privileged basis does not exist.
Nevertheless, one can always find an almost selfdual basis and one 
can proceed much along the same lines with minor modifications (the
interested reader can consult the comprehensive review~\cite{Bjork:2008}
for a full account of these methods).

It is interesting to stress that for $n$ qubits, the reference
state (\ref{Psi_n2}) can be elegantly written in terms of the field
elements $\Gal{2^{n}}$ as follows
\begin{equation}
  | \Psi_{0} \rangle = \frac{1}{( 1 +  \xi^{2} )^{n/2}}
  \sum_{\alpha \in \Gal{2^{n}}} \xi^{h ( \alpha )} \, | \alpha \rangle \, ,
  \label{gen_eq}
\end{equation}
where the function $h ( \alpha )$ counts the number of nonzero
coefficients $a_{j}$ in the expansion of $\alpha$ in the 
basis.

The operator transforming from an arbitrary basis 
$\{ \theta^{\prime}_{j} \}$ into the selfdual one $\{ \theta_{j} \}$  
is given by
\begin{equation}
  \op{\mathcal{T}} = \sum_{\mu \in GF(2^{n})} 
   | m_{1}, \ldots, m_{n} \rangle 
   \langle m_{1}^{\prime}, \ldots, m_{n}^{\prime} |  \, ,  
  \label{T}
\end{equation}
where
\begin{equation}
  \mu = \sum_{j} m_{j}^{\prime} \theta_{j}^{\prime}  =
  \sum_{j} m_{j} \theta_{j}  \, . 
\end{equation}
The operator $\op{\mathcal{T}} $ is always a permutation and  plays 
the role of an entangling (nonlocal) operator.

Let us examine the simple yet illustrative example of a two-qubit
coherent state. According to Eq.~(\ref{gen_eq}), we have
\begin{equation}
  | \Psi_{0} \rangle = \frac{1}{1 +  \xi^{2}} 
  ( | 0 \rangle + \xi | \sigma \rangle + \xi | \sigma^{2} \rangle +
  \xi^{2} | \sigma^{3} \rangle ) \, ,
  \label{2q_cs}
\end{equation}
where $\sigma$ is a primitive element. The selfdual basis is 
$\{ \sigma , \sigma^{2} \}$, and we have the representation
\begin{eqnarray}
  |0 \rangle = |0 0 \rangle =  
  \left ( 
    \begin{array}{c}
      0 \\ 
      0 \\ 
      0 \\ 
      1
    \end{array}
  \right ) \, ,
  \qquad
  | \sigma \rangle = |1 0 \rangle = 
  \left ( 
    \begin{array}{c}
      0 \\ 
      0 \\ 
      1 \\ 
      0
    \end{array}
  \right ) \, , & \nonumber \\
  \\
  | \sigma^{2} \rangle = |0 1 \rangle =
  \left ( 
    \begin{array}{c}
      0 \\ 
      1 \\ 
      0 \\ 
      0
    \end{array}
  \right ) \, , 
  \qquad
  | \sigma^{3} \rangle = |1 1 \rangle =
  \left ( 
    \begin{array}{c}
      1 \\ 
      0 \\ 
      0 \\ 
      0
    \end{array}
  \right ) \, . \nonumber  
  \label{2q_basis}
\end{eqnarray}
In consequence,
\begin{eqnarray}
  | \Psi_{0} \rangle & = & 
  \frac{1}{1 + \xi^{2}}
  \left ( 
    \begin{array}{c}
      \xi ^{2} \\ 
      \xi \\ 
      \xi \\ 
      1
    \end{array}
  \right ) \nonumber \\
  &  =  &
  \frac{1}{\sqrt{1 + \xi^{2}}}
  \left ( 
    \begin{array}{c}
      \xi \\ 
      1
    \end{array}
  \right ) 
  \otimes 
  \frac{1}{\sqrt{1 + \xi^{2}}}
  \left ( 
    \begin{array}{c}
      \xi \\ 
      1
    \end{array}
  \right ) \, .
\end{eqnarray}

In a non selfdual basis, such as $\{ \sigma , \sigma^{3} \}$, we have
\begin{equation}
| 0 \rangle =  |0 0 \rangle \, , \quad
| \sigma \rangle  = | 1 0 \rangle \, , \quad 
| \sigma^{3} \rangle = | 0  1\rangle \, , \quad
| \sigma^{2}\rangle = | 1 1 \rangle \, ,
\end{equation}
and 
\begin{equation}
| \Psi_0 \rangle = \frac{1}{1 + \xi^{2}}
\left ( 
\begin{array}{c}
\xi \\ 
\xi^{2} \\ 
\xi \\ 
1
\end{array}
\right ) \, .
\end{equation}
The transition operator (\ref{T}) turns out to be
\begin{equation}
\op{\mathcal{T}} = 
\left ( 
\begin{array}{cccc}
0 & 1 & 0 & 0 \\ 
1 & 0 & 0 & 0 \\ 
0 & 0 & 1 & 0 \\ 
0 & 0 & 0 & 1
\end{array}
\right )  \, ,
\end{equation}
which is nothing but a matrix representation of the CNOT operator.

\subsection{Quasidistribution functions}

The displacement operators (\ref{eq:phiodd}) immediately suggest to
introduce an $s$-ordered kernel
\begin{eqnarray}
  \op{w}^{(s)} (\mu , \nu ) & = & \frac{1}{d^{n}} 
  \sum_{\lambda ,\kappa}\chi (\mu \lambda - \nu \kappa ) \, \op{D}(\mu ,\nu ) 
  \nonumber \\
  & \times & \langle \Psi_{0} | \op{D} (\mu, \nu ) | \Psi_{0} \rangle^{-s} \, ,
\end{eqnarray}
which, in view of Eq.~(\ref{FTcomp}), can also be interpreted as a
double Fourier transform of $\op{D}(\mu ,\nu )$. We can next introduce $s$-ordered quasidistribution functions through
\begin{equation}
  W^{(s)}_{\op{\varrho}} (\mu , \nu )= \Tr [ \op{\varrho}\,
  \op{w}^{(s)} (\mu , \nu )] \, ,  
  \label{eq:Wigdn}
\end{equation}
and the inversion relation reads as
\begin{equation}
  \op{\varrho} = \frac{1}{d^{n}} \sum_{\mu ,\nu} 
  \op{w}^{(-s)}(\mu ,\nu ) \, W^{(s)}_{\op{\varrho}}(\mu , \nu ) \, .
\end{equation}

Due to the factorization of the character in the selfdual basis, 
the kernels $ \op{w}^{(s)} (\mu, \nu)$ factorize in this basis
\begin{equation}
  \op{w}^{(s)} ( \mu ,\nu ) = \prod_{j} \op{w}^{(s)} ( m_{j}, n_{j}) \, ,
\end{equation}
and, consequently, also do the corresponding quasidistributions
\begin{equation}
  W_{\op{\varrho}}^{(s)} ( \mu , \nu ) = \prod_{j} 
  W_{\op{\varrho}_{j}}^{(s)} ( m_{j}, n_{j} ) \, .
\end{equation}

For the particular case of the Wigner function, one can check that
\begin{equation}
  \sum_{\mu ,\nu} W_{\op{\varrho}} (\mu ,\nu ) \,\delta_{\nu ,
    \alpha \mu + \beta} = \sum_{\mu ,\nu} W_{\op{\varrho}}(\mu ,\nu ) \,
  \delta_{\nu ,-\alpha^{-1}\mu
    -\beta} \, ,
\end{equation}
that is, the sum over a line of slope $\alpha $ is the same as over a
line of slope $ - \alpha^{-1}$. The sum over the axes $\mu $\ and $\nu
$\ are thus equal
\begin{equation}
  \sum_{\mu ,\nu} W_{\op{\varrho}} (\mu ,\nu ) \,\delta_{\nu ,0} = 
    \sum_{\mu , \nu} W_{\op{\varrho}} (\mu ,\nu ) \,\delta_{\mu ,0} \, .
\end{equation}

\begin{figure}
  \includegraphics[width=0.85\columnwidth]{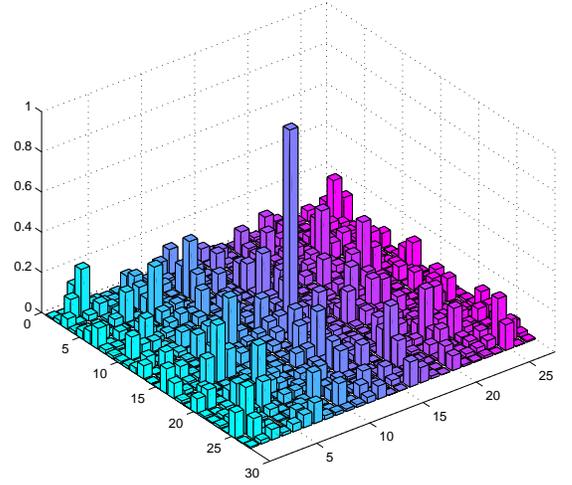}
  \caption{$Q$ function for the reference state $| \Psi_{0} \rangle$, 
 for a system of three qutrits. The order in the axes is as
follows: $\sigma^{13}$, $\sigma^{17}$, $\sigma^{14}$, $\sigma$,
$\sigma^{2}$, $\sigma^{21}$, $\sigma^{23}$, $\sigma^{7}$, $\sigma^{15}$,
$\sigma^{4}$, $\sigma^{16}$, $\sigma^{6}$, $\sigma^{8}$, 0,
$\sigma^{9}$, $\sigma^{12}$, $\sigma^{25}$, $\sigma^{24}$,
$\sigma^{5}$, $\sigma^{3}$, $\sigma^{19}$, $\sigma^{11}$,
$\sigma^{22}$, $\sigma^{10}$, $\sigma^{20}$, $\sigma^{18}$,
$\sigma^{26}$, with $\sigma$ the primitive element.}
  \label{Fig2}
\end{figure}

Note also, that the $Q$ function reduces to
\begin{equation}
  \label{eq:2}
  Q_{\op{\varrho}} (\mu , \nu ) = 
  \langle \mu ,\nu |\op{\varrho} | \mu ,\nu \rangle \, .
\end{equation}
In Fig.~\ref{Fig2} we have plotted this $Q$ function for the reference
state $|\Psi_{0} \rangle$ in a system of three qutrits. The 
selfdual basis here is $\{ \sigma, \sigma^{3}, \sigma^{9} \}$
and the primitive element is a solution of the irreducible 
polynomial  $x^{3} + 2 x^{2} + 1 = 0$.

%%%%%%%%%%%%%%%%%%%%%%%%%%%%%%%%%%%%%%%%%%%%%%%%%
%%%%%%%%%%%%%%%%%%%%%%%%%%%%%%%%%%%%%%%%%%%%%%%%

\section{Squeezed states}

Squeezed states constitute a simple nontrivial enlargement of the
notion of coherent states. In continuous variables, a squeezed state
is a minimum uncertainty state that my have less fluctuations in one
quadrature ($\op{q}$ or $\op{p}$) than a coherent state. They are
generated from the vacuum by using the unitary squeeze operator
\begin{equation}
  \op{S} ( \mathfrak{s} ) = \exp [ - i \mathfrak{s} \, 
  (\op{q} \op{p} + \op{p} \op{q} ) ] \, ,
\end{equation}
with a subsequent displacement to an arbitrary point in the complex
plane
\begin{equation}
  \label{eq:SSgen}
  | q, p; \mathfrak{s} \rangle = \op{D} (q, p) \op{S} (\mathfrak{s} ) \,
  | \psi_{0} \rangle \, .
\end{equation} 
It is easy to check that
\begin{equation}
  \label{eq:sqact}
  \op{S} (\mathfrak{s}) \, \op{q} \, \op{S}^{\dagger} (\mathfrak{s} ) = 
  \op{q} \, e^{\mathfrak{s}} \, , 
  \qquad
  \op{S} (\mathfrak{s}) \, \op{p} \, \op{S}^{\dagger} (\mathfrak{s} ) = 
  \op{p} \, e^{- \mathfrak{s}} \, ,
\end{equation}
so that, the operator $\op{S} (\mathfrak{s})$ attenuates one
quadrature and amplifies the canonical one by the same factor
determined by the squeeze factor $\mathfrak{s}$, which, for
simplicity, we have taken as real. As a simple consequence of
(\ref{eq:sqact}) one can verify the transformations for $\op{U} (q)$
and $\op{V} (p)$:
\begin{equation}
  \op{S} (\mathfrak{s} ) \, \op{U} (q) \, \op{S}^{\dagger} (\mathfrak{s} ) 
  = U^{\mathfrak{s}} (q) \, , 
  \quad 
  \op{S} (\mathfrak{s} ) \, \op{V} (p) \, \op{S}^{\dagger} (\mathfrak{s} )
  = \op{V}^{- \mathfrak{s}} (p) \, . 
  \label{SUV}
\end{equation}

For a single qudit, squeezed states have been recently considered in
detail in Ref.~\cite{Marchiolli:2007}, using an extended
Cahill-Glauber formalism. Here, we prefer to follow an alternative
approach and define a squeeze operator as
\begin{equation}
  \op{S}_{s} = \sum_{\ell} | \ell \rangle \langle s \ell | \, ,
  \qquad  
  s \in \mathbb{Z}_{d} \, .
  \label{S_d}
\end{equation}
At first sight, this can appear as a rather abstract choice. However,
notice that
\begin{equation}
  \op{S}_{s}^{\dagger} \, \op{U}^{n} \, \op{S}^{\dagger}_{s} = 
  \op{U}^{n \, s} \, , 
  \qquad 
  \op{S}_{s} \, \op{V}^{m} \, \op{S}_{s}  = \op{V}^{m \, s^{-1}} \, ,  
  \label{UVS_d}
\end{equation}
which is a direct translation of the action (\ref{SUV}) to this
discrete case. This also means that in the squeezed ``vacuum''
\begin{equation}
  \label{eq:sqvac}
  | \psi_{0}; s \rangle = \op{S}_{s} | \psi_{0} \rangle \, ,
\end{equation}
the average values of some powers of the displacement operators are
the same
\begin{equation}
  \langle \psi_{0}; s | \op{U} | \psi_{0}; s \rangle = 
  \langle \psi_{0}; s | \op{V}^{s^{2}} | \psi_{0}; s \rangle \, .  
  \label{psi_S_d}
\end{equation}

Perhaps, the clearest way to visualize this squeezing is to use a
quasidistribution, such as, e.g., the Wigner function. If 
$\op{\varrho}_{s} = \op{S}_{s} \, \op{\varrho} \, \op{S}_{s}^{\dagger}$ 
denotes the density operator of a squeezed state, we have
\begin{equation}
  W_{\op{\varrho}_{s}} ( m, n ) = W_{\op{\varrho}} ( s m, s^{-1} n ) \, ,
  \label{W_S_d}
\end{equation}
whose geometrical interpretation is obvious and is the phase-space 
counterpart of the property (\ref{eq:sqact}). For reasons
that will become evident soon, we refer to this as ``geometrical
squeezing''.  We also note the following symmetry property of the
Wigner function
\begin{equation}
  \sum_{m,n} W_{\op{\varrho}_{s}} ( m, n ) \, \delta_{n, 0} =
  \sum_{m,n} W_{\op{\varrho}_{s^{-1}}} ( m, n ) \,  \delta_{m,0} \, .
\end{equation}

\begin{figure}
  \includegraphics[width=0.85\columnwidth]{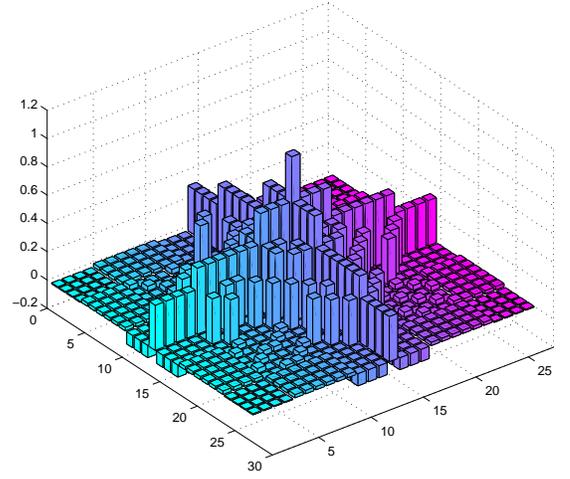}
  \caption{Wigner function for a squeezed ``vacuum'' state $|
    \Psi_{0}, \varsigma \rangle$, for a system of three qutrits, 
    with the same order in the axes as in Fig.~2.}
  \label{Fig3}
\end{figure}

For many qudits, our developed intuition suggests a direct translation
of (\ref{S_d}) in terms of the field elements in $\Gal{d^{n}}$, namely
\begin{equation}
  \op{S}_{\varsigma} = \sum_{\lambda} |\lambda \rangle 
  \langle \varsigma \lambda | \, ,
  \qquad
  \varsigma \in \Gal{d^{n}} \, ,
  \label{S_dn}
\end{equation}
in terms of which we can write relations similar to
Eqs.(\ref{UVS_d})-(\ref{W_S_d}).  In fact, one can define a squeezed
``vacumm'' as in Eq.~(\ref{eq:sqvac}), i.e., $ | \Psi_{0}; \varsigma
\rangle = \op{S}_{\varsigma} | \Psi_{0} \rangle$. In Fig.~\ref{Fig3}
we plot the Wigner function for this squeezed state in a system of
three qutrits with $\varsigma = \sigma^{7}$.

%%%%%%%%%%%%%%%%%%%%%%%%%%%%%%%%%%%%%%%%%%%%%%%%%
%%%%%%%%%%%%%%%%%%%%%%%%%%%%%%%%%%%%%%%%%%%%%%%%

Nevertheless, now the squeezing acquires a new physical perspective:
the squeeze operator (\ref{S_dn}) cannot be, in general, factorized
into a product of single qudit squeezing operators. This means that by
applying $\op{S}_{\varsigma}$ to a factorized state we generate
correlations between qudits; i.e., we create entangled states.  The
most striking example is of course the $n$ qubit case, since there is
no single qubit squeezing.

To understand these correlations consider a general factorized state
\begin{equation}
  | \Psi \rangle = 
  \sum_{\lambda} \! C_{\lambda} | \lambda \rangle = 
  \sum_{c_{\ell_{1}}, \ldots, c_{\ell_{n}}} \! \! \!
  c_{\ell_{1}} \ldots c_{\ell_{n}} \, | \ell_{1}, \ldots, \ell_{n} \rangle \, ,  \label{psi_S}
\end{equation}
and  apply (\ref{S_dn}).  The resulting state turns out to be
\begin{equation}
  \op{S}_{\varsigma} | \Psi \rangle = \sum_{\ell_{1}, \cdots, \ell_{n}} 
  C_{m_{1} \theta_{1} + \ldots + m_{n} \theta_{n}}  \, 
  | \ell_{1}, \ldots,  \ell_{n} \rangle \, ,
\end{equation}
where
\begin{eqnarray}
  & \displaystyle
  m_{i} = \sum_{j,k=0}^{d-1} f_{ijk} \ell_{j} h_{k} \, , & \nonumber \\
  & & \\ 
  & f_{ijk}= \tr ( \theta_{i} \theta_{j} \theta_{k}) \, , 
  \quad 
  h_{k}= \tr (\varsigma^{-1} \theta_{k}) \, , & \nonumber 
\end{eqnarray}
$\tr$ (written in lower case) is the trace operation in the field
(see the Appendix) and $\{ \theta_{j} \}$ is the basis.

As a example let us consider a three-qubit system. Now the selfdual
basis is $\{ \theta_{1}= \sigma^{3}, \theta_{2} = \sigma^{5},
\theta_{3} = \sigma^{6}\}$, where $\sigma $ is a primitive element,
solution of the irreducible polynomial $x^{3} + x + 1 = 0$.  The
result of applying $\op{S}_{\sigma^{k}}$ to the state (\ref{psi_S})
can be expressed in terms of
\begin{eqnarray}
  \op{S}_{\sigma} | \Psi \rangle & = & 
  \sum_{\lambda \in \Gal{2^{3}}} \! \!
  C_{\sigma^{6} \lambda} | \lambda \rangle = 
  \sum_{p, q, r \in \mathbb{Z}_{2}} c_{p+q} c_{p+r} c_{q} 
  |p, q, r \rangle \, ,
  \nonumber \\
 & & \\
  \op{S}_{\sigma^{3}} | \Psi \rangle & = & 
\sum_{\lambda \in \Gal{2^{3}}}  \! \!
C_{\sigma^{4} \lambda} | \lambda \rangle =   
\sum_{p, q, r \in \mathbb{Z}_{2}} c_{p+q+r} c_{p+r} c_{r}
|p, q, r \rangle \,  .  \nonumber
\end{eqnarray}
In fact, the transformations $\{ \op{S}_{\sigma^{5}}, 
\op{S}_{\sigma^{6}}\}$ generate the same entanglement (except
for permutations) as $\op{S}_{\sigma^{3}}$, while $\{ \op{S}_{\theta^{2}}, \op{S}_{\theta^{4}} \}$ generate the same entanglement (again except for permutations) as $\op{S}_{\sigma}$.

\section{Concluding remarks}

In summary, we have provided a handy toolbox for dealing with
many-qudit systems in phase space. The mathematical basis of our
approach is the use algebraic field extensions that produce results 
in composite dimensions in a manner very close to the continuous
case.

Another major advantage of our theory relies on the use of the
finite Fourier transform and its eigenstates for the definition of
coherent states. We believe that this makes a clear connection with
the standard coherent states for continuous variables and constitutes
an elegant solution to this problem. The factorization properties of 
the resulting coherent states in different bases is also an interesting
question.

We have also established a set of important results that have allowed
us to obtain discrete analogs of squeezed states. While for a single 
qudit, these squeezed states have the properties one would expect from 
our continuous-variable experience, for many qudits an amazing relation 
with entanglement appears.

We think that the techniques presented here are more than a mere
academic curiosity, for they are immediately applicable to a variety
of experiments involving qudit systems.

\appendix

\section{Finite fields}
\label{Sec: Galois}

In this appendix we briefly recall the minimum background needed in
this paper.  The reader interested in more mathematical details is
referred, e.g., to the excellent monograph by Lidl and
Niederreiter~\cite{Lidl:1986}.

A commutative ring is a nonempty set $R$ furnished with two binary
operations, called addition and multiplication, such that it is an
Abelian group with respect the addition, and the multiplication is
associative.  Perhaps, the motivating example is the ring of integers
$\mathbb{Z}$, with the standard sum and multiplication. On the other
hand, the simplest example of a finite ring is the set $\mathbb{Z}_n$
of integers modulo $n$, which has exactly $n$ elements.

A field $F$ is a commutative ring with division, that is, such that 0
does not equal 1 and all elements of $F$ except 0 have a
multiplicative inverse (note that 0 and 1 here stand for the identity
elements for the addition and multiplication, respectively, which may
differ from the familiar real numbers 0 and 1). Elements of a field
form Abelian groups with respect to addition and multiplication (in
this latter case, the zero element is excluded).

The characteristic of a finite field is the smallest integer $d$ such
that
\begin{equation}
  d \, 1= \underbrace{1 + 1 + \ldots + 1}_{\mbox{\scriptsize $d$ times}}=0
\end{equation}
and it is always a prime number. Any finite field contains a prime
subfield $\mathbb{Z}_d$ and has $d^n$ elements, where $n$ is a natural
number. Moreover, the finite field containing $d^{n}$ elements is
unique and is called the Galois field $\Gal{d^n}$.

Let us denote as $\mathbb{Z}_{d} [x]$ the ring of polynomials with
coefficients in $\mathbb{Z}_{d}$. Let $P(x)$ be an irreducible
polynomial of degree $n$ (i.e., one that cannot be factorized over
$\mathbb{Z}_{d}$). Then, the quotient space $\mathbb{Z}_{d}[X]/P(x)$
provides an adequate representation of $\Gal{d^n}$. Its elements can
be written as polynomials that are defined modulo the irreducible
polynomial $P(x)$. The multiplicative group of $\Gal{d^n}$ is cyclic
and its generator is called a primitive element of the field.

As a simple example of a nonprime field, we consider the polynomial
$x^{2}+x+1=0$, which is irreducible in $\mathbb{Z}_{2}$.  If $\prim$
is a root of this polynomial, the elements $\{ 0, 1, \prim , \prim^{2}
= \prim + 1 = \prim^{-1} \} $ form the finite field $\Gal{2^2}$ and
$\prim$ is a primitive element.

A basic map is the trace
\begin{equation}
  \label{deftr}
  \tr (\lambda ) = \lambda + \lambda^{2} + \ldots +
  \lambda^{d^{n-1}} \, .
\end{equation}
It is always in the prime field $\mathbb{Z}_d$ and satisfies
\begin{equation}
  \label{tracesum}
  \tr ( \lambda + \lambda^\prime ) =
  \tr ( \lambda ) + \tr ( \lambda^\prime ) \, .
\end{equation}
In terms of it we define the additive characters as
\begin{equation}
  \label{Eq: addchardef}
  \chi (\lambda ) = \exp \left [ \frac{2 \pi i}{p}
    \tr ( \lambda ) \right] \, ,
\end{equation}
which posses two important properties:
\begin{equation}
  \chi (\lambda + \lambda^\prime ) =
  \chi (\lambda ) \chi ( \lambda^\prime ) ,
  \qquad
  \sum_{\lambda^\prime \in
    \Gal{d^n}} \chi ( \lambda \lambda^\prime ) = d^n
  \delta_{0,\lambda} \, .
  \label{eq:addcharprop}
\end{equation}

Any finite field $\Gal{d^n}$ can be also considered as an
$n$-dimensional linear vector space. Given a basis $\{ \theta_{j} \}$,
($j = 1,\ldots, n$) in this vector space, any field element can be
represented as
\begin{equation}
  \label{mapnum}
  \lambda = \sum_{j=1}^{n} \ell_{j} \, \theta_{j} ,
\end{equation}
with $\ell_{j}\in \mathbb{Z}_{d}$. In this way, we map each element of
$\Gal{d^n}$ onto an ordered set of natural numbers $\lambda
\Leftrightarrow (\ell_{1}, \ldots , \ell_{n})$.

Two bases $\{ \theta_{1}, \ldots, \theta_{n} \} $ and $\{
\theta_{1}^\prime, \ldots , \theta_{n}^\prime \} $ are dual when
\begin{equation}
  \tr ( \theta_{k} \theta_{l}^\prime ) =\delta_{k,l}.
\end{equation}
A basis that is dual to itself is called selfdual.

There are several natural bases in $\Gal{d^n}$. One is the polynomial
basis, defined as
\begin{equation}
  \label{polynomial}
  \{1, \prim, \prim^{2}, \ldots, \prim^{n-1} \} ,
\end{equation}
where $\prim $ is a primitive element. An alternative is the normal
basis, constituted of
\begin{equation}
  \label{normal}
  \{\prim, \prim^{d}, \ldots, \prim^{d^{n-1}} \}.
\end{equation}
The choice of the appropriate basis depends on the specific problem at
hand. For example, in $\Gal{2^2}$ the elements $\{ \prim ,
\prim^{2}\}$ are both roots of the irreducible polynomial. The
polynomial basis is $\{ 1, \prim \} $ and its dual is $\{ \prim^{2}, 1
\}$, while the normal basis $\{ \prim , \prim^{2} \} $ is selfdual.
  
The selfdual basis exists if and only if either $d$ is even or both
$n$ and $d$ are odd. However for every prime power $d^n$, there exists
an almost selfdual basis of $\Gal{d^n}$, which satisfies the
properties: $\tr ( \theta_{i} \theta_{j} ) =0$ when $i\neq j$ and $\tr
( \theta_{i}^{2} ) =1$, with one possible exception. For instance, in
the case of two qutrits $\Gal{3^2}$, a selfdual basis does not exist
and two elements $\{ \prim^2, \prim^4\}$, $\prim$ being a root of the
irreducible polynomial $x^2 + x + 2 =0$, form an almost selfdual basis
\begin{equation}
  \label{eq:aqpp}
  \tr (\prim^2 \prim^2) = 1 \, , \quad
  \tr (\prim^4 \prim^4) = 2 \, , \quad 
  \tr (\prim^2 \prim^4) = 0 \, . 
\end{equation}

%\bibliography{FiniteQM}

\end{document}